\documentclass[amssymb,longbibliography,notitlepage,superscriptaddress]{revtex4-1}
\usepackage[utf8]{inputenc}
\usepackage{graphicx}
\usepackage{float}
\usepackage{amsmath}
\usepackage{subcaption}

\begin{document}

\title{Locating nuclear-powered submarines with antineutrinos}

\author{Dr. Sven-Patrik Hallsj\"o}
\email{patrik.hallsjo@gmail.com}
\affiliation{Independent researcher, Stockholm, Sweden}


\date{April 2026}

    \begin{abstract}
    Nuclear-powered submarines are difficult to track with conventional methods in congested waterways. We revisit antineutrino-based detection as a barrier concept, analogous to a neutrino-enabled SOSUS-style fence in strategic straits. Using analytic scaling relations and numerical estimates, we show that detectability depends primarily on closest approach, detector depth, and deployed mass. For representative assumptions, a 20\,kt detector in the Strait of Gibraltar reaches a local benchmark score $Z_A\simeq2.54$ for an assumed 100\,MW thermal-power sensitivity-study case in a conservative worst-case transit (with Poisson operating point $(P_\mathrm{FA},P_\mathrm{det})\simeq(5.5\times10^{-3},0.51)$ at threshold $k=2$), while a three-detector line raises the mapped score to $Z_A\simeq4.66$. For broad ocean passages such as GIUK, required detector counts are substantially larger; in the baseline maximum passing distance $\mathrm{PDD}_{\max}=5$\,km geometry, about 80 detectors yield only $Z_A\sim1.6$. The paper outlines detector technology choices, statistical assumptions, and deployment constraints for a first-generation feasibility assessment.
    \end{abstract}
 
\maketitle

\section{Introduction}

Nuclear energy has revolutionized submarine warfare. Military
submarines can stay submerged for months and move stealthily at high
speeds through the oceans.  Submarine-launched ballistic missiles
(SLBM) are the backbone of nuclear deterrence since they offer an
assured retaliatory capability. A submarine whose location has been
uncovered by an adversary is usually relatively defenseless and thus
avoiding detection is \emph{the} operational priority of any submarine
captain. Any major breakthrough in either avoiding detection or
improving detection would have major consequences for naval and
nuclear doctrine.

Conventional means to detect submarines center on acoustic signatures,
but also electromagnetic, temperature, chemical signatures and many
others have been proposed, including neutrinos~\cite{ciaasw}.

During the Cold War the focus was on finding or concealing SLBM
carrying submarines in the open, deep blue ocean and a detailed
analysis of neutrino detection can be found in a JASON
report~\cite{JASON}. The conclusion is rather negative, however the
basic framework remains sound. Neutrino
detector technology has however made great progress in the past 30
years. Common detector types for charged current neutrino reactions
are either liquid scintillator, {\it e.g.}
KamLAND~\cite{Eguchi:2002dm}, or water Cherenkov types, {\it e.g.}
Super-Kamiokande~\cite{Fukuda:1998mi}. Both detector types profit from
the addition of gadolinium to improve neutron
tagging~\cite{An:2012eh,Beacom:2003nk}. Also neutral current reactions
of low-energy neutrinos have been measured; for a recent result see
Ref.~\cite{Akimov:2017ade}.

In view of changed geopolitics, notably with China as major
emerging naval force and India fielding SLBMs and North Korea at least
exploring the concept, and advances in neutrino detector technology
we will revisit the question of how to detect nuclear-powered
submarines using their neutrino signatures.

\section{Neutrino detection}
\label{sec:detection}

Neutrinos were discovered in 1956 by Cowan and Reines using the
P-reactor at Savannah River as a source~\cite{Cowan:1992xc}. They are
electrically neutral, nearly massless, spin-1/2 particles. In this
paper, we are exclusively concerned with electron antineutrinos,
which we refer to as neutrinos for simplicity.
In nuclear fission, neutrinos result from the beta decays of neutron
rich fission fragments and on average about 2 detectable neutrinos per
fission are produced with a mean energy of around
$4\,\mathrm{MeV}$. For neutrinos of this energy the most practical
detection reaction is inverse beta decay (IBD) on free protons
$\bar\nu_e + p \rightarrow e^+ + n$ with a cross section of around
$10^{-42}\,\mathrm{cm}^2$. This small cross section is the reason that
neutrinos are \emph{not} attenuated by seawater (or any other
material).

\begin{figure}
  \includegraphics[width=\textwidth]{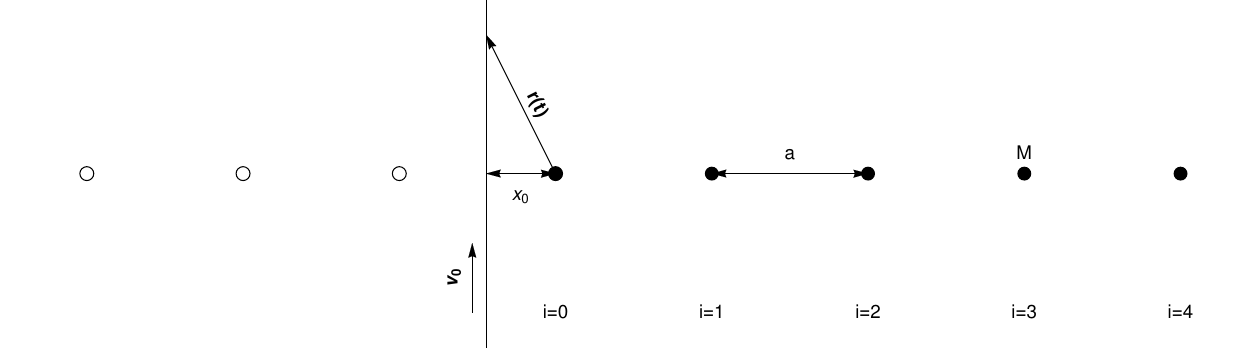}
  \caption{\label{fig:ske} Schematic of detector arrangement and
    submarine course: The vertical line represents the course of the
    submarine and the black circles (open and filled) each represent a
    neutrino detector of mass $M$.}
\end{figure}

We first consider simple cases that are amenable to analytical
calculation before turning to full numerical analysis. Consider a
single detector of mass $M$ and a submarine cruising on a straight
course at constant depth $d$ with constant speed $v_0$. A simple
propulsion scaling assumes that the reactor power follows, as an
idealized first-order model~\cite{JASON},

\begin{equation}
P=P_\mathrm{max}\left(\frac{v_0}{v_\mathrm{max}}\right)^3\,,
\end{equation}
which corresponds to the standard turbulent-drag scaling argument
$D\propto v^2$ with propulsive power $P_\mathrm{prop}=Dv$. Real naval
propulsion can deviate from this idealized limit because of
speed-dependent propulsor efficiency, cavitation constraints, and
non-propulsive hotel loads. In the uncertainty treatment below, we
therefore interpret the cubic law as a reference model and treat the
effective exponent as a systematic range
$P\propto v^n$ with $2\lesssim n\lesssim 4$, where $P_\mathrm{max}$ and
$v_\mathrm{max}$ are the maximum rated
power and speed of the submarine. The neutrino signal rate is then
given by~\cite{Formaggio:2013kya,Adey:2019ywk,Kopeikin:2004cn}:
\begin{equation}
    s(t)=\underbrace{\sigma_\mathrm{IBD}N_A f_p M_D \frac{P_\mathrm{max}}{4\pi E_\mathrm{fission}}}_{=:N_0}\left(\frac{v_0}{v_\mathrm{max}}\right)^3\frac{1}{r(t)^2}\,
\end{equation}
where $\sigma_\mathrm{IBD}=6.10\times10^{-43}\,\mathrm{cm}^2$ is the
inverse beta-decay cross section per fission of
uranium-235~\cite{Adey:2019ywk}, $N_A$ is Avogadro's constant, $f_p$
is the number of protons per gram of detector material (for water,
$f_p\simeq 2/18$), $M_D$ is the detector mass in grams, and
$E_\mathrm{fission}=201.92\,\mathrm{MeV}$ is the thermal energy released
by one fission in uranium-235~\cite{Kopeikin:2004cn}. The quantity
$r(t)=|\mathbf{r}(t)|$ denotes the distance between detector and
submarine. If $x_0$ denotes the closest approach of the submarine to
the detector (Fig.~\ref{fig:ske}), then
\begin{equation}
\label{eq:r}
    r^2(t)=x_0^2+v_0^2t^2\,.
\end{equation}
The duration $t_{1/2}$ over which the signal exceeds half of its
maximum value is given by $t_{1/2}=2 x_0/v_0$. For integrated-yield
expressions below, we define $N:=N_0/v_\mathrm{max}$.
In reality the IBD cross section is energy dependent
($\sigma_\mathrm{IBD}(E_\nu)\propto E_ep_e$), and neutrino oscillations
at km--100 km baselines can induce normalization shifts at the
$\mathcal{O}(10\%)$ level. In this first-pass study we use a
spectrum-averaged effective cross section and absorb these effects into
the quoted systematic envelope.

The total time-integrated number of signal events is given by
\begin{equation}
\label{eq:signal_infinite}
S=\int_{-\infty}^{+\infty}\,s(t)\,dt=N\frac{\pi}{x_0}\left(\frac{v_0}{v_\mathrm{max}}\right)^2=0.309\,\left(\frac{P_\mathrm{max}}{100\,\mathrm{MW}}\right)\left(\frac{20\,\mathrm{kn}}{v_\mathrm{max}}\right)\left(\frac{v_0}{v_\mathrm{max}}\right)^2\left(\frac{1\,\mathrm{km}}{x_0}\right)\left(\frac{M}{1\,\mathrm{kt}}\right)\,.
\end{equation}
Hence, for the cubic propulsion model, the integrated signal scales as
$S\propto(v_0/v_\mathrm{max})^2$: slow transits reduce signal yield.
The normalization still depends on $P_\mathrm{max}/v_\mathrm{max}$,
which is a measure of how effectively nuclear power is converted to
propulsion. For liquid scintillator instead of water, the prefactor is
0.397. Successful detection requires sufficiently small $x_0$, either
through geographic constraints ({\it e.g.}, the Strait of Gibraltar)
or through a line of detectors with spacing $a$; in the latter case,
$x_0\leq a/2$ if detector and submarine are at the same depth $d$.
During $t_{1/2}$, approximately $S/2$ events are registered. For
$P_\mathrm{max}=100\,\mathrm{MW}$, $v_\mathrm{max}=20\,\mathrm{kn}$,
$x_0=1\,\mathrm{km}$, and detector mass $20$\,kt, we obtain $S=6.2$;
thus $S/2=3.1$ events occur in less than
$x_0/v_0\simeq 100$\,s when $v_0=v_\mathrm{max}$.

\begin{figure}[h]
    \centering
    \includegraphics[width=0.85\textwidth]{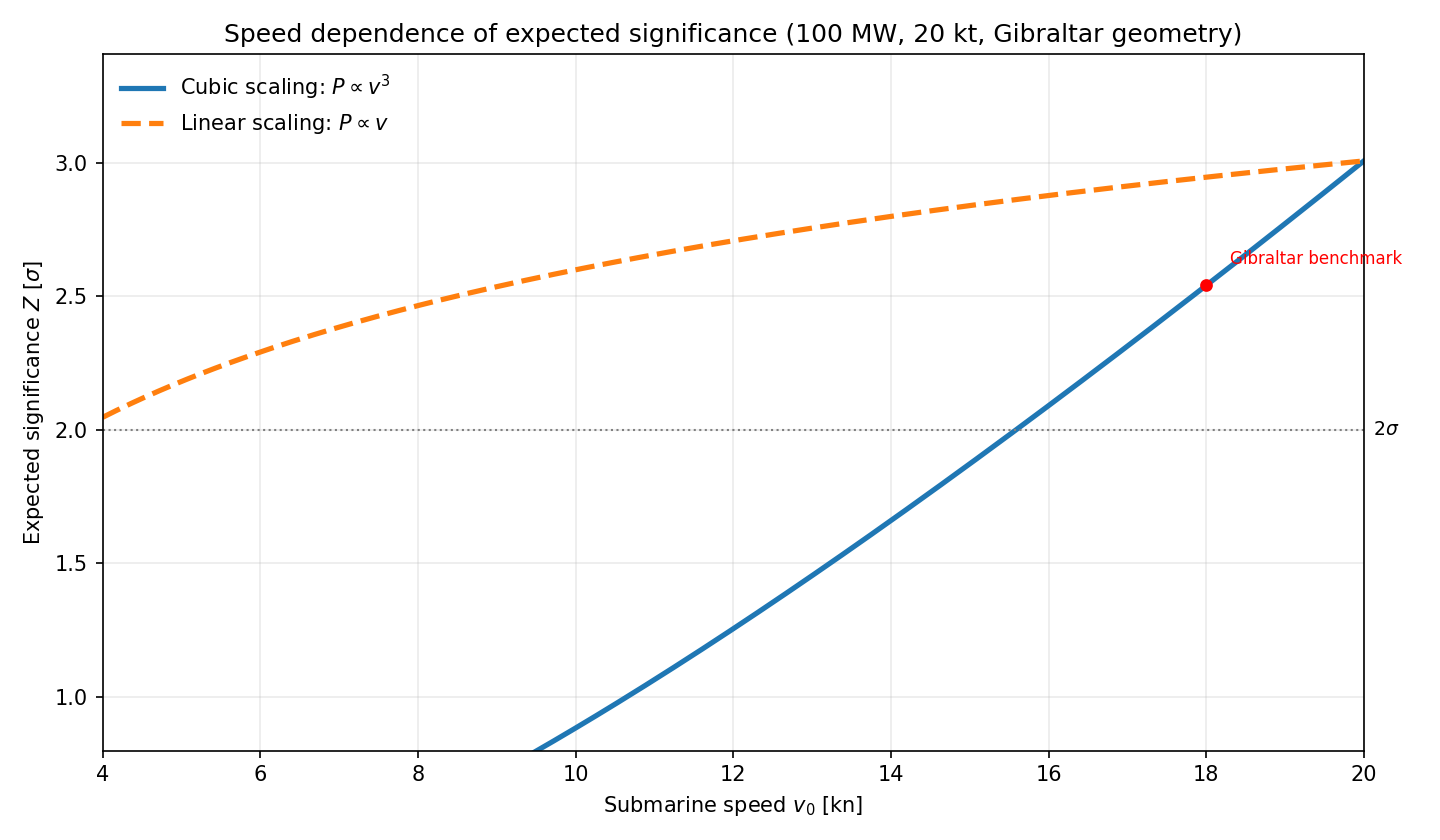}
    \caption{Expected significance versus submarine speed for two
    propulsion-power scalings in Gibraltar-like geometry
    ($\mathrm{PDD}_{\max}=3.04$\,km, depth 300\,m, detector mass 20\,kt,
    nominal reactor scale 100\,MW).}
    \label{fig:SpeedDependence}
\end{figure}

Figure~\ref{fig:SpeedDependence} illustrates the dependence of expected
significance on submarine speed for two propulsion assumptions. For the
cubic law $P\propto v^3$, slow transits strongly suppress the signal.
For a linear law $P\propto v$, the speed dependence is much weaker,
because the signal remains approximately constant while integrated
background increases for slower transits.

\begin{figure}[h]
    \centering
    \includegraphics[width=\textwidth]{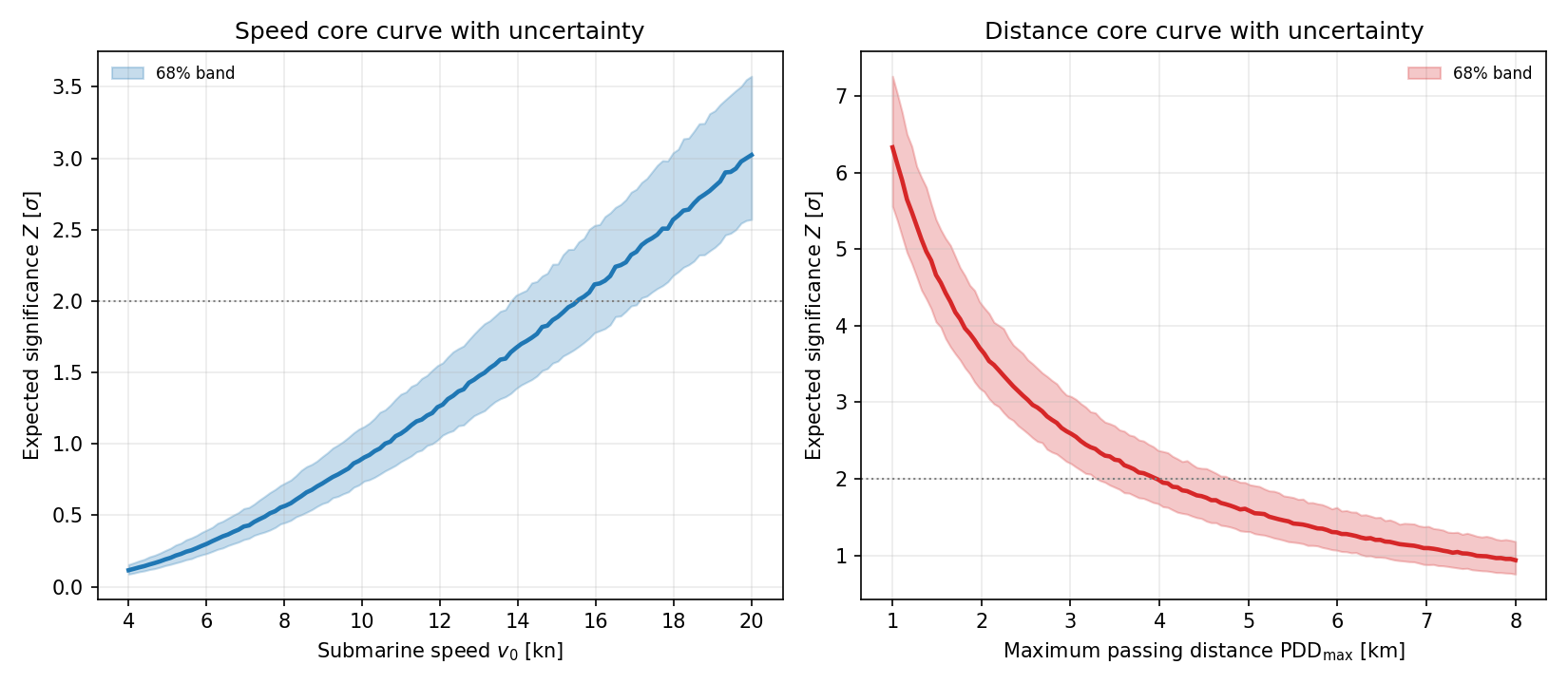}
    \caption{Uncertainty bands on two core curves obtained from
    Monte Carlo propagation of background, efficiency, and trajectory
    uncertainties.}
    \label{fig:CoreUncertaintyBands}
\end{figure}

To quantify modeling uncertainty on core trends, we propagate
three nuisance terms: a 30\% background normalization uncertainty,
a 10\% detection-efficiency uncertainty, and a 15\% trajectory-distance
uncertainty. Figure~\ref{fig:CoreUncertaintyBands} shows the resulting
68\% bands for the speed and distance core curves.

At the Gibraltar benchmark point, this propagation gives
$Z_{16\%}=2.17$, $Z_{50\%}=2.56$, and $Z_{84\%}=3.04$.

In an infinite chain of detectors, one also has an infinite detector
mass, and hence $S$ diverges. However, we are interested in the
statistical significance of a detection, which for a single detector
can be approximated as~\cite{Cowan:2010js}
\begin{equation}
    \chi^2_1=\frac{S^2}{S+B}\,,
\end{equation}
where $B$ is the total number of background events observed. This
$\chi^2$ form is strictly accurate only for counting experiments with
$S,B>10$, but it is still useful for understanding limiting cases.
In this work, this $\chi^2$ subsection is used for asymptotic scaling
intuition only; quantitative sensitivities and threshold statements are
based on the Asimov counting significance in
Sec.~\ref{subsec:stats}.
Assuming that the background is the same for each detector and
replacing $x_0$ in Eq.~\ref{eq:r} with $a/2+ia$, where $i$ is the
index numbering detectors, one finds that $\chi^2$ for a semi-infinite
array, $i=0,1,\ldots,\infty$ is given by
\begin{equation}
  \label{eq:array}
    \chi^2_\infty=\frac{N}{a}\left[\gamma + \ln 4 +\psi^{(0)}\!\left(\frac{1+\kappa}{2\kappa}\right)\right]\,,
\end{equation}
where $\psi^{(0)}(z)$ is the digamma function~\cite{AbramowitzStegun1964} and $\gamma$ is Euler's
constant. Let
$B:=\kappa S_0$, with $S_0$
being the signal in the detector closest to the track and $\kappa$ is the
inverse signal-to-background ratio for that detector.

In the limit $\kappa\rightarrow0$, this function diverges, as does the
total signal rate. In the opposite limit,
$\kappa\rightarrow\infty$, it goes to zero; in that limit, the ratio
$\chi^2_\infty/\chi^2_1$ approaches $\pi^2/8\simeq 1.23$. For
$\kappa=1$, $\chi^2_\infty/\chi^2_1=\ln 4 \simeq 1.39$, and we see
that the array gain remains relatively small, with most of the effect
coming from the handful of detectors closest to the course of the
submarine; in fact, the 4 closest detectors yield already 85\% of a
full array. One can also see from Eq.~\ref{eq:array} that distributing
total detector mass along a given segment of the array into many
smaller detectors and thus reducing the spacing $a$ will not change
$\chi^2_\infty$ in the case that there are no backgrounds,
$\kappa\rightarrow0$. For the more realistic case of backgrounds,
reducing $a$ will decrease $x_0$ and thus decrease $\kappa$. Indeed
for large backgrounds, the best solution would be to have one long
(and narrow) continuous detector.

Using IBD as the detection reaction, one can employ delayed
coincidence between the prompt energy deposited by the positron and
the later ($\sim100\,\mu\mathrm{s}$) energy release from neutron
capture. This double signature effectively suppresses most
backgrounds, especially those from natural radioactivity in detector
materials. The remaining backgrounds stem from cosmic rays and are
therefore most effectively reduced by increasing detector depth.
Following detailed experimental studies for reactor
neutrinos~\cite{Abe:2012ar}, we retain the depth dependence but
calibrate the overall normalization to the Gibraltar benchmark used in
this study. The background rate is parameterized as
\begin{equation}
  b(d,M)=b_\mathrm{ref}\left(\frac{d}{300\,\mathrm{m}}\right)^{-1.705}\left(\frac{M}{20\,\mathrm{kt}}\right)\,,
  \end{equation}
with $b_\mathrm{ref}=3.03\times10^{-4}\,\mathrm{min}^{-1}$.
Event counts are obtained from this rate via the analysis time window,
e.g. $B=b\,\Delta t$ in one bin or
$B_\mathrm{tot}=\int b\,dt$ over a transit.
Operationally, $b$ is an effective sum of components,
$b=b_\mu+b_\mathrm{spall}+b_\mathrm{geo}+b_\mathrm{reactor}+b_\mathrm{inst}$,
which are not individually resolved in the present baseline model.
For scale anchoring, this effective normalization implies about
$b_\mathrm{ref}\times 1440\simeq0.44$ events/day at 300\,m depth for a
20\,kt detector, and about $0.10$ events/day at a JUNO-like depth
($\sim700$\,m) under the same scaling law. These are intended as
order-of-magnitude, post-selection effective rates compatible with
reactor-neutrino low-background practice (e.g. KamLAND/JUNO and
Gd-tagged water-Cherenkov programs) rather than detector-specific
predictions~\cite{Eguchi:2002dm,75Juno,Beacom:2003nk}.

\begin{figure}[h]
    \centering
    \begin{subfigure}[b]{0.3\textwidth}
        \includegraphics[width=\textwidth]{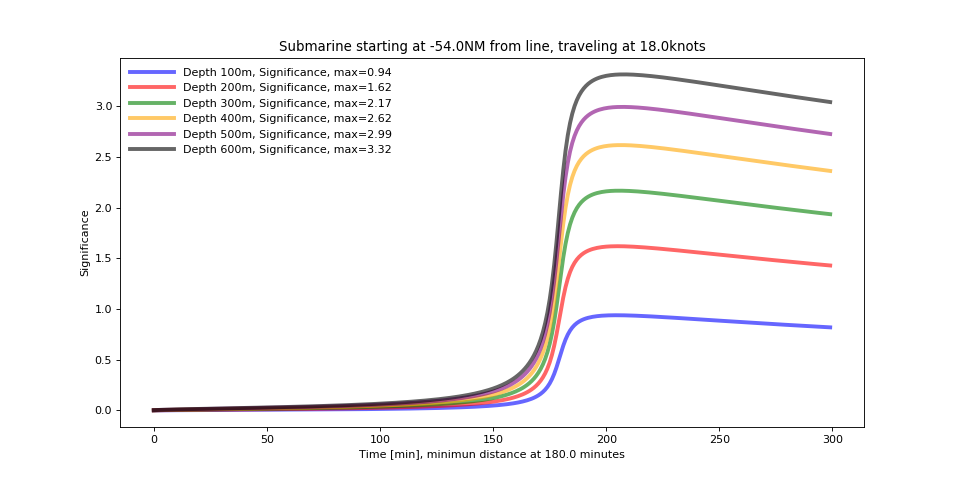}
        \caption{Varying detector depth, 5km separation, 100 MW, 20 kt.}
        \label{fig:VaryingDepth}
    \end{subfigure}
    ~
    \begin{subfigure}[b]{0.3\textwidth}
        \includegraphics[width=\textwidth]{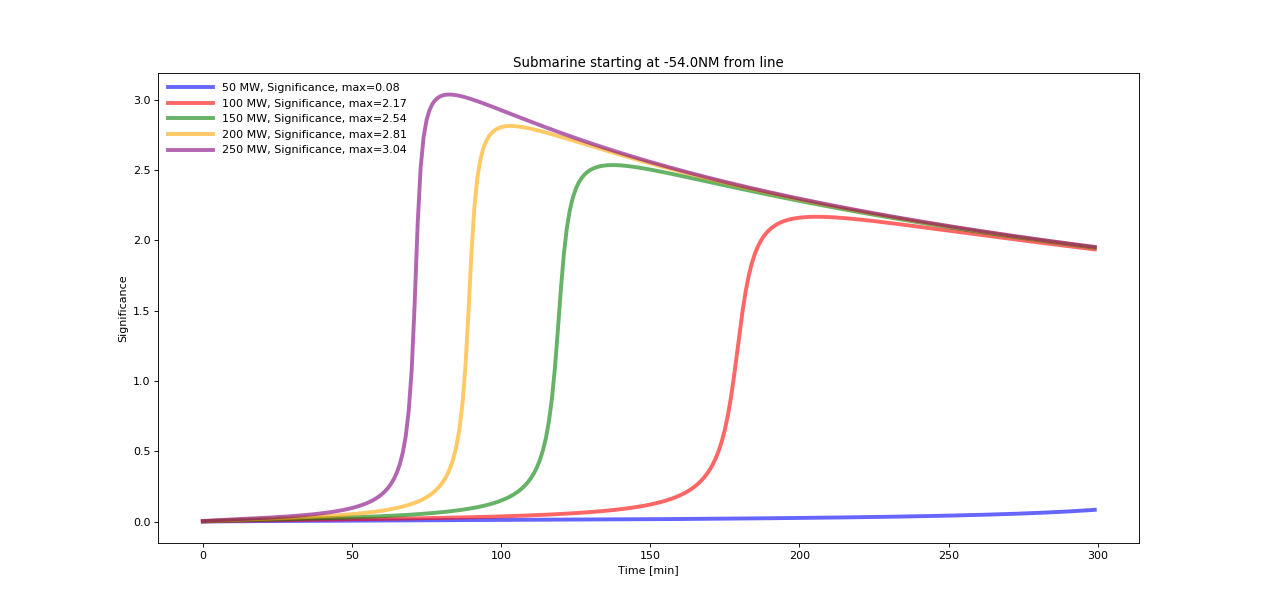}
        \caption{Vary reactor power, 5km separation 300m depth, 20 kt.}
        \label{fig:ReactorPower}
    \end{subfigure}
    \begin{subfigure}[b]{0.3\textwidth}
        \includegraphics[width=\textwidth]{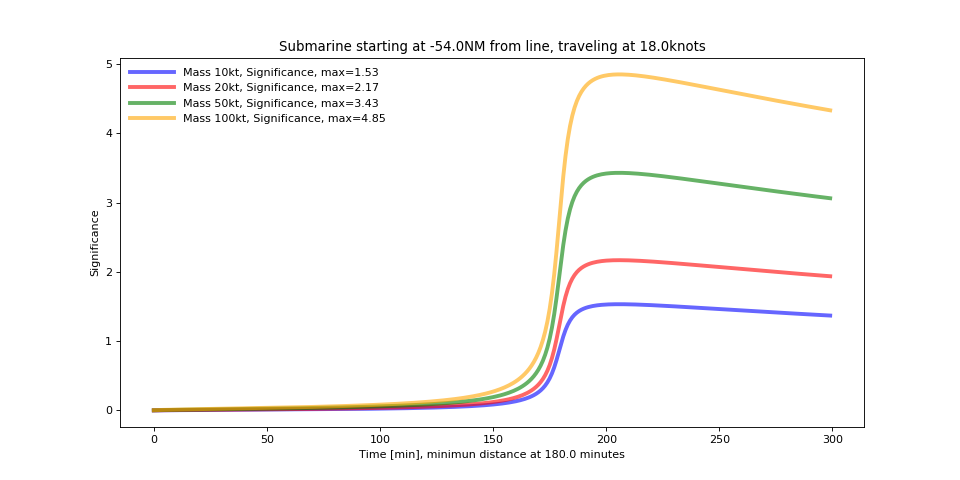}
        \caption{Vary detector mass, 5km separation, 300m depth, 100 MW. }
        \label{fig:DetectorMass}
    \end{subfigure}
    \caption{Variations of the different factors}\label{fig:VDF}
\end{figure}

Figure~\ref{fig:VDF} illustrates the dominant trends used throughout
this study: increasing depth and detector mass improve significance,
while higher reactor power extends the range for a fixed significance
threshold.

\subsection{From signal and background to significance}
\label{subsec:signalbg}

For a finite transit window of duration $T$, the expected signal count
is
\begin{equation}
S(T)=\int_{-T/2}^{+T/2} s(t)\,dt
=N\left(\frac{v_0}{v_\mathrm{max}}\right)^2\frac{2}{x_0}
\arctan\left(\frac{v_0T}{2x_0}\right)\,.
\label{eq:signal_window}
\end{equation}
The infinite-window limit of Eq.~\ref{eq:signal_window} reproduces the
compact scaling in Eq.~\ref{eq:signal_infinite}.

Assuming approximately constant background rate over the same window,
the expected background count is
\begin{equation}
B(T)=\int_{-T/2}^{+T/2} b\,dt\simeq b(d,M)\,T\,.
\label{eq:background_window}
\end{equation}

For compact cross-scenario visualization, we also report an
Asimov-equivalent local score~\cite{Cowan:2010js}
\begin{equation}
Z_A=\sqrt{2\left[(S+B)\ln\left(1+\frac{S}{B}\right)-S\right]}\,.
\label{eq:asimovZ}
\end{equation}

For the benchmark configuration in the Strait of Gibraltar
($P_\mathrm{max}=100$\,MW, $M=20$\,kt, $x_0=3.041$\,km,
$v_0=18$\,kn, $T=360$\,min), this pipeline gives
$S\simeq1.61$ events, $B\simeq0.109$ events, and
$Z_A\simeq2.51$, consistent with the stated Gibraltar benchmark
within rounding and window-approximation differences. In this
low-count regime, $Z_A$ is used as a descriptive mapping, whereas
operational decisions are based on Poisson operating points.

\begin{figure}[h]
    \centering
    \includegraphics[width=0.95\textwidth]{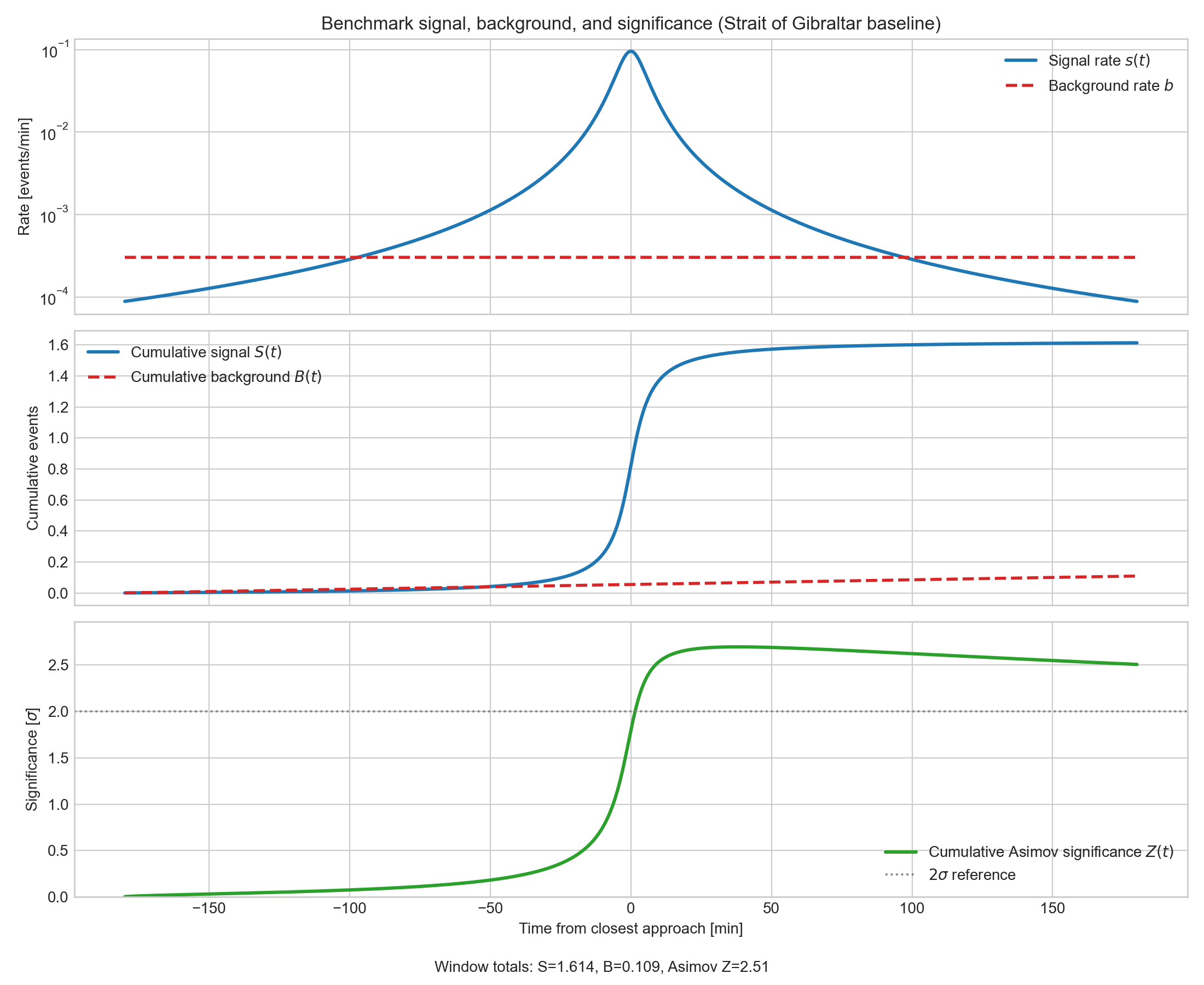}
    \caption{Section-2 workflow illustration for a Gibraltar baseline:
    top panel shows signal and background rates versus time from closest
    approach; middle panel shows cumulative signal and background counts;
    bottom panel shows cumulative Asimov significance $Z(t)$ with a
    $2\sigma$ reference line.}
    \label{fig:SignalVsBackgroundProfile}
\end{figure}

\section{Detector choices}

The preceding theoretical discussion established that, under suitable
geometric and depth conditions, event rates can be sufficient for a
timely detection of fast-moving submarines. This requires detectors in
the 1--100\,kt mass range that can withstand deep-sea operation and
remain autonomous for extended periods.

There are two mature media for reactor-antineutrino detection at this
scale: liquid organic scintillator (LS) and gadolinium-doped water
(GW). LS is derived from mineral oil and doped to scintillate,
converting ionization into light at roughly 10,000 photons per MeV.
This relatively high light yield enables detection of neutron capture
on hydrogen (2.2\,MeV gamma cascade). In water, light is produced by
the Cherenkov effect with much lower yield (roughly 100 photons per
MeV), so gadolinium loading (typically $\sim0.1\%$) is needed to obtain
a bright delayed neutron-capture signature (about 8\,MeV in gamma rays).

\begin{itemize}
\item acoustic visibility and density matching to seawater,
\item buoyancy control and long-term mooring stability,
\item filtration and radiopurity control in the target medium,
\item photosensor coverage, timing resolution, and calibration.
\end{itemize}

In both media, the physical signal is a faint prompt-delayed light
pair detected by photomultiplier tubes (PMTs). For this first
feasibility study we adopt an LS baseline comparable to JUNO in mass
scale ($\sim20$\,kt)~\cite{75Juno}. Depth remains the dominant
background-control parameter: for representative configurations, moving
from 300\,m depth to 100\,m can reduce significance to around
$1.6\sigma$.

All quoted depths assume that the full active detector volume is below
the stated overburden; this makes the background estimate slightly
conservative.

\subsection{Statistical treatment}
\label{subsec:stats}

The simulation uses 1-minute time bins. Signal and background counts
are treated as Poisson variables, and the primary performance metrics
are $P_\mathrm{FA}(k)$ and $P_\mathrm{det}(k)$ from Poisson tails.
The Asimov score in Eq.~\ref{eq:asimovZ} is retained as a secondary,
smooth descriptor to compare scenarios and draw contour plots.

Unless explicitly stated otherwise,
results below are quoted for a single detector; combining neighboring
detectors would increase sensitivity. Quoted thresholds are local
(single-hypothesis) values;
global false-alert rates under continuous multi-trajectory scanning are
left for dedicated operational studies.

To make the low-count regime explicit, we also evaluate count-threshold
tests directly with Poisson tails~\cite{Cowan:2010js}:
\begin{equation}
P_\mathrm{FA}(k)=\sum_{n=k}^{\infty}\mathrm{Pois}(n\mid B)\,,\qquad
P_\mathrm{det}(k)=\sum_{n=k}^{\infty}\mathrm{Pois}(n\mid S+B)\,.
\end{equation}
For the Gibraltar benchmark ($S\simeq1.61$, $B\simeq0.109$), one finds
$(P_\mathrm{FA},P_\mathrm{det})\simeq(0.103,0.821)$ for $k=1$,
$(5.53\times10^{-3},0.514)$ for $k=2$, and
$(1.99\times10^{-4},0.249)$ for $k=3$.
With $N_\mathrm{trial}$ independent tests per day, the global false-alert
probability is approximately
$p_\mathrm{global}\simeq1-(1-p_\mathrm{local})^{N_\mathrm{trial}}\simeq N_\mathrm{trial}p_\mathrm{local}$,
so even modest local $p_\mathrm{local}$ values can imply high operational
false-alert rates~\cite{Gross:2010qma}.

\begin{figure}[h]
    \centering
    \includegraphics[width=0.95\textwidth]{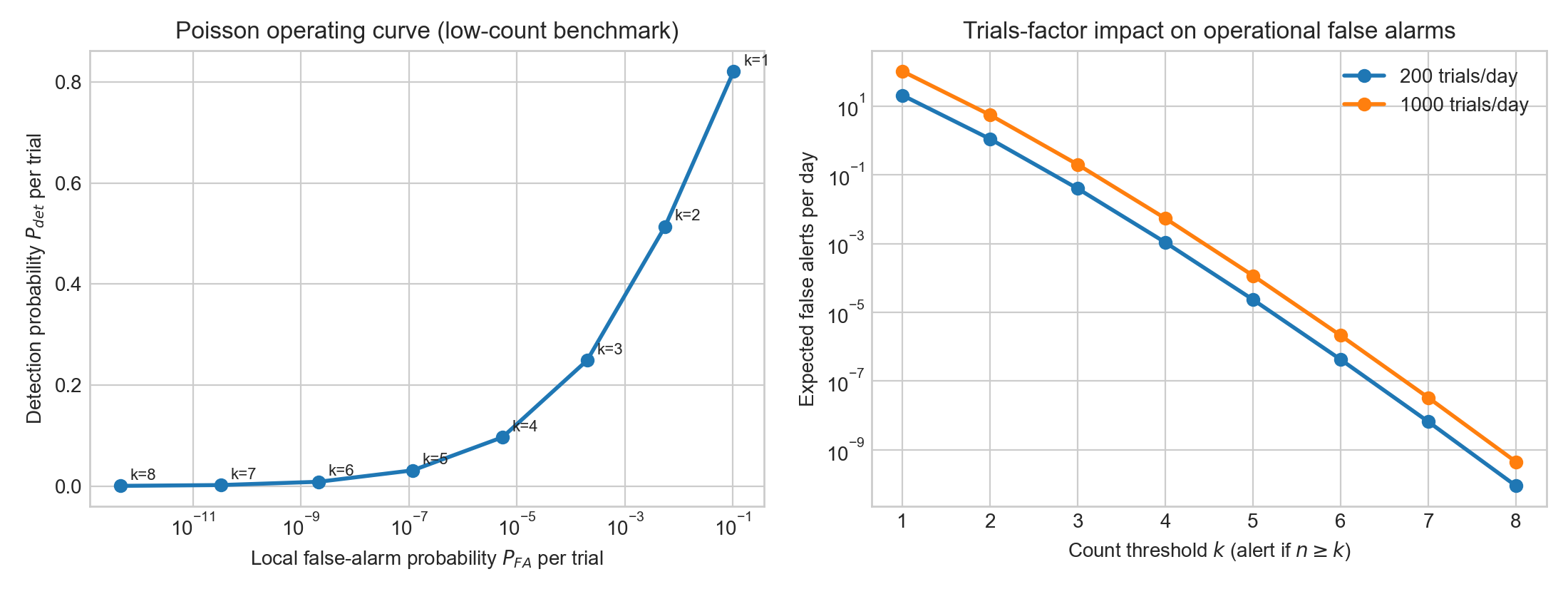}
    \caption{Poisson-count operating characteristics for the Gibraltar
    low-count benchmark. Left: per-trial detection probability versus
    false-alarm probability for count thresholds $k$. Right: expected
    false alerts per day after applying a trials factor.}
    \label{fig:PoissonOperating}
\end{figure}

\subsection{Reproducibility inputs}

Table~\ref{tab:repro-inputs} consolidates the baseline numerical inputs
used in the benchmark scenarios and figures.

\begin{table}[h]
\centering
\begin{tabular}{ll}
\hline
Quantity & Baseline assumption \\
\hline
Detector medium & Liquid scintillator (LS) \\
Fiducial detector mass & $M=20$\,kt \\
Reference transit speed & $v_0=18$\,kn \\
Rated speed (normalization) & $v_\mathrm{max}=20$\,kn \\
Transit window & $\pm54$\,NM around closest approach \\
Time bin size & $\Delta t=1$\,min \\
Propulsion exponent range & $2\leq n\leq4$ (reference: $n=3$) \\
Background depth scaling & $b\propto d^{-1.705}$ \\
Background calibration at reference point & $b_\mathrm{ref}=3.03\times10^{-4}$ events/min \\
Reference transit duration & $T_\mathrm{ref}=360$\,min \\
Reference background count & $B_\mathrm{ref}=0.109$ events/transit \\
Primary operating metric & Poisson $(P_\mathrm{FA},P_\mathrm{det})$ versus count threshold $k$ \\
Secondary descriptor & Asimov-equivalent local score $Z_A$ \\
Reference local alert rule & $k=2$ count threshold (benchmark) \\
Nuisance widths & $\sigma_b/b=30\%$, $\sigma_\epsilon/\epsilon=10\%$, $\sigma_x/x=15\%$ \\
\hline
\end{tabular}
\caption{Consolidated baseline inputs used for the main numerical
scenarios in this feasibility study.}
\label{tab:repro-inputs}
\end{table}

\section{Potential locations}
From the depth-significance trends in Fig.~\ref{fig:VDF}, depths of
order 300\,m or greater are preferred for practical deployments. We now
apply the model to a small subset of representative straits.

We define $\mathrm{PDD}_{\max}$ as the maximum passing distance to the
detector under the assumed route geometry.

\begin{figure}[h]
    \centering
    \includegraphics[width=0.9\textwidth]{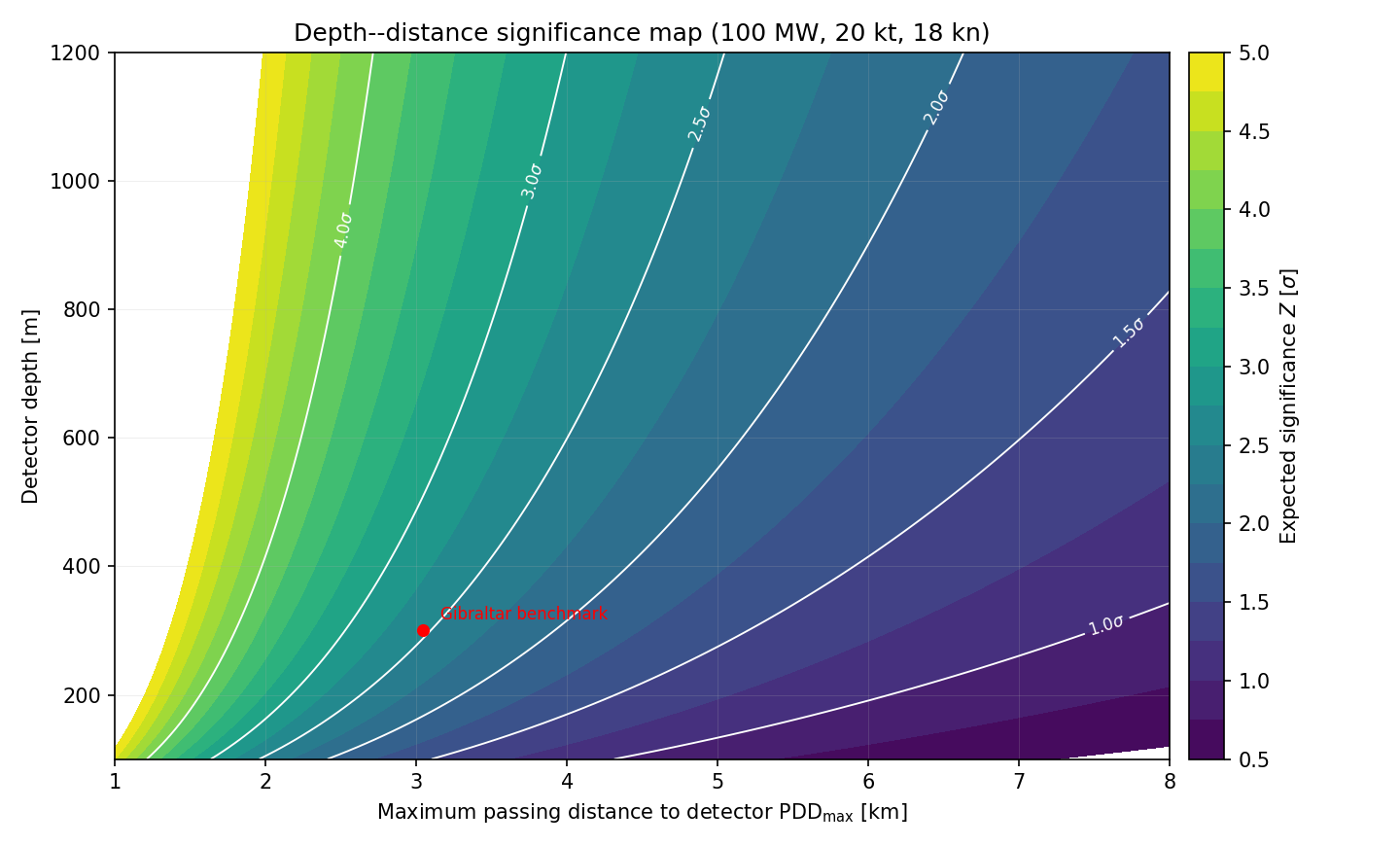}
    \caption{Depth--distance significance map for a baseline
    100\,MW submarine and a 20\,kt detector. The red marker denotes the
    Gibraltar benchmark used for normalization.}
    \label{fig:DepthPDDContour}
\end{figure}

As an immediate design aid, Fig.~\ref{fig:DepthPDDContour} shows a
depth--distance significance map for a baseline case
($P=100$\,MW, $M=20$\,kt, $v_0=18$\,kn). The map uses the same
Asimov-equivalent descriptor $Z_A$ as Sec.~\ref{subsec:stats} and the
same depth-scaling
exponent for background; the overall background normalization is
calibrated to the Gibraltar benchmark point.

\begin{table}[h]
\centering
\begin{tabular}{lcc}
\hline
Case & Configuration & Asimov-equivalent score $Z_A$ \\
\hline
Gibraltar (single) & 1 detector, $P=100$\,MW, $\mathrm{PDD}_{\max}=3.04$\,km & $2.54$ \\
Gibraltar (line) & 3 detectors near coast, $P=100$\,MW & $4.66$ \\
GIUK (Gibraltar-like PDD) & $\sim133$ detectors for comparable geometry & $\sim2.5$ \\
GIUK ($\mathrm{PDD}_{\max}=5$\,km) & $\sim80$ detectors & $1.6$ \\
\hline
\end{tabular}
\caption{Representative scenarios used in this study.}
\label{tab:location-summary}
\end{table}

\subsection{Gibraltar}\label{subsec:Gibraltar}

\begin{figure}[h]
    \centering
    \begin{subfigure}[b]{0.4\textwidth}
        \includegraphics[width=\textwidth]{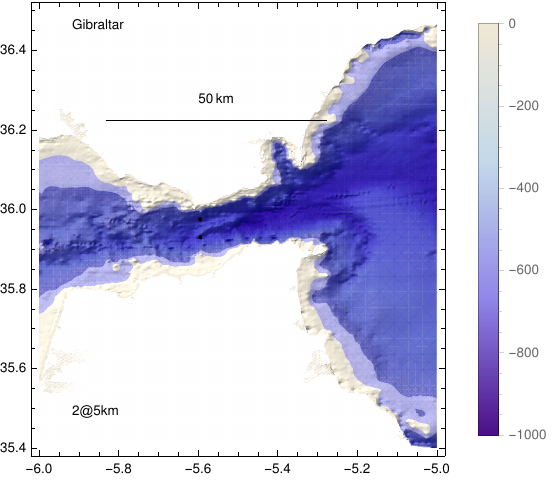}
        \caption{A map of the Strait of Gibraltar.}
        \label{fig:GibraltarMap}
    \end{subfigure}
    ~
    \begin{subfigure}[b]{0.4\textwidth}
        \includegraphics[width=\textwidth]{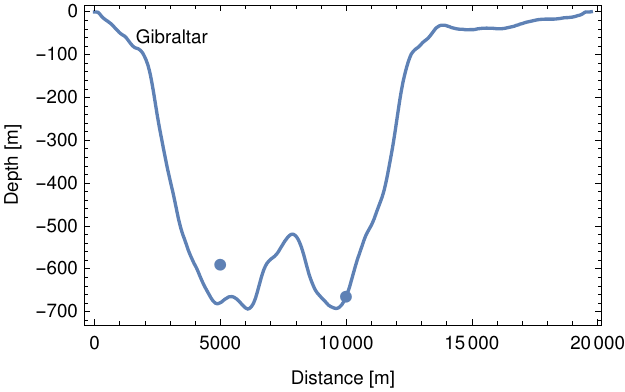}
        \caption{A depth profile of the Strait of Gibraltar.}
        \label{fig:GibraltarProfile}
    \end{subfigure}
    \caption{The Strait of Gibraltar.}\label{fig:Gibraltar}
\end{figure}

For this study we use the heavily trafficked strait between Spain and
North Africa. As shown in Fig.~\ref{fig:GibraltarProfile}, local
bathymetry offers plausible detector siting points.

\begin{figure}[h]
    \centering
    \includegraphics[width=\textwidth]{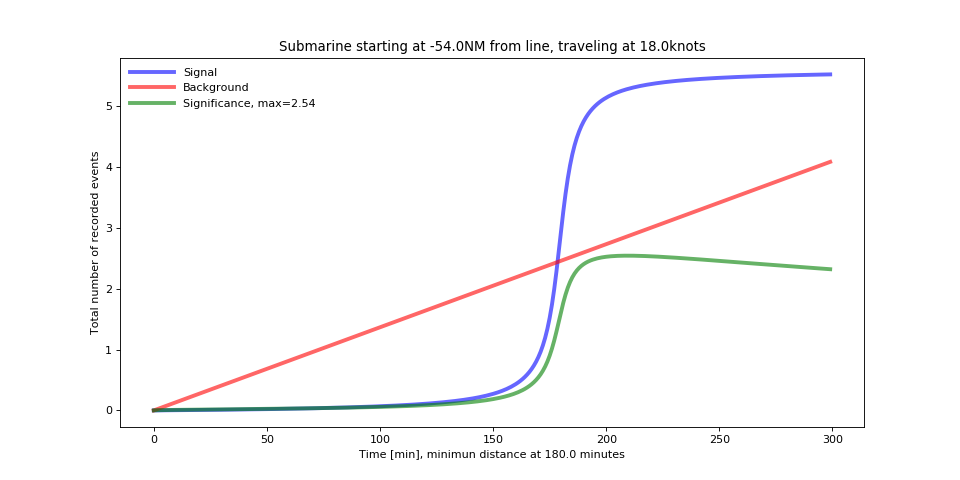}
    \caption{The Gibraltar study}
    \label{fig:GibraltarStudy}
\end{figure}

As a first approximation, we model the strait as a 6\,km wide and
500\,m deep rectangle. With one detector centered in the strait, the
maximum passing distance to the detector is
$\mathrm{PDD}_{\max}=\sqrt{(6000/2)^2+500^2}\simeq3041$\,m.
Assuming a 100\,MW submarine moving at 18\,kn on a straight transit
(start/end 54\,NM from the detector), the scenario yields an
Asimov-equivalent score of $Z_A\simeq2.54$ in
Fig.~\ref{fig:GibraltarStudy}; for this same benchmark, the Poisson
operating point at $k=2$ is
$(P_\mathrm{FA},P_\mathrm{det})\simeq(5.5\times10^{-3},0.51)$.
Raising reactor power to 200\,MW increases the mapped score to about
$Z_A\simeq3.2$.

Keeping reactor power at 100\,MW but deploying a 3-detector line near
the coasts (approximately halving $\mathrm{PDD}_{\max}$) increases the
mapped score to $Z_A\simeq4.66$. This remains a conservative worst-case
transit where the submarine maximizes its distance to known detector
locations.

\subsection{Greenland--Iceland--UK (GIUK gap)}

\begin{figure}[h]
    \centering
    \includegraphics[width=0.7\textwidth]{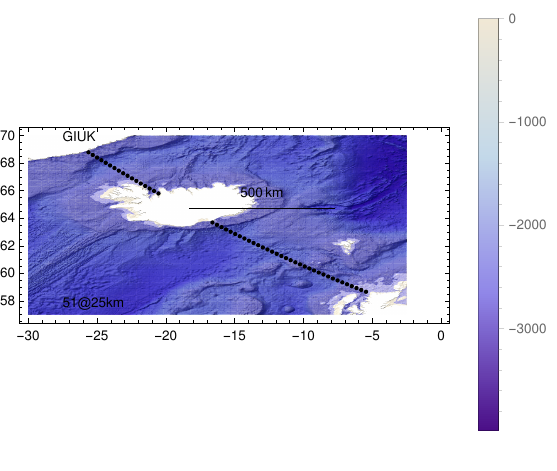}
    \caption{Map overview of the GIUK gap.}\label{fig:GIUK}
\end{figure}

The GIUK gap is historically strategic and much wider than Gibraltar.
Approximating it as an $\sim800$\,km wide and 500\,m deep corridor,
and targeting Gibraltar-like $\mathrm{PDD}_{\max}$ performance,
requires about 133 detectors. Relaxing to
$\mathrm{PDD}_{\max}=5$\,km reduces the requirement to roughly
80 detectors, with significance dropping to about $1.6\sigma$.

This comparison highlights a central conclusion: neutrino fencing is
most practical in geographically constrained chokepoints and becomes
rapidly resource-intensive for open-ocean barriers.

\subsection{Model scope and limitations}

The current framework is deliberately a first-pass feasibility model.
Three simplifications are especially important when interpreting
headline numbers: (i) trajectories are idealized as straight,
constant-speed transits rather than adversarially optimized routes;
(ii) the background model is an effective depth-scaled surrogate rather
than a fully component-resolved detector simulation; and (iii)
systematics are represented by independent nuisance widths, without a
full correlation model or profile-likelihood/Bayesian marginalization
over all operational parameters. A production-grade assessment should
therefore combine trajectory Monte Carlo, component-level backgrounds,
and continuously scanned (global) false-alarm accounting.

\section{Conclusions}
This feasibility study indicates that antineutrino-based submarine
detection is physically plausible in narrow, sufficiently deep straits.
For realistic detector masses ($\mathcal{O}(10)$\,kt), significance is
driven by closest approach distance and overburden depth, with
multi-detector layouts providing significant operational margin.

The Gibraltar case demonstrates that conservative transit assumptions
can still yield detection-level significance with modest deployments,
while the GIUK case shows that scaling to wide barriers rapidly becomes
cost-dominant in this baseline model. Tsushima and Danish-strait
scenarios appear intermediate, but require route-specific quantification
before firm module-count conclusions. A neutrino fence should therefore
be viewed as a specialized complement to conventional systems (e.g.,
acoustic monitoring), not as a universal replacement. The quoted
sensitivities should be interpreted as upper-bound performance under
non-adversarial transit assumptions.

\bibliography{references}

\newpage

\appendix

\section{Alternative Locations}

\subsection{The Danish Straits}

\begin{figure}[h]
    \centering
    \begin{subfigure}[b]{0.4\textwidth}
        \includegraphics[width=\textwidth]{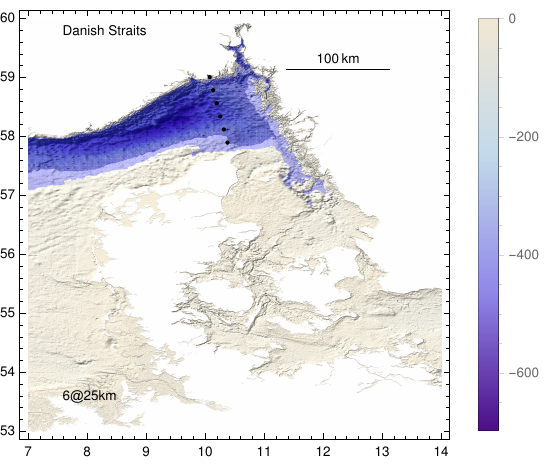}
        \caption{A map of the Danish strait.}
        \label{fig:DanishMap}
    \end{subfigure}
    ~
    \begin{subfigure}[b]{0.4\textwidth}
        \includegraphics[width=\textwidth]{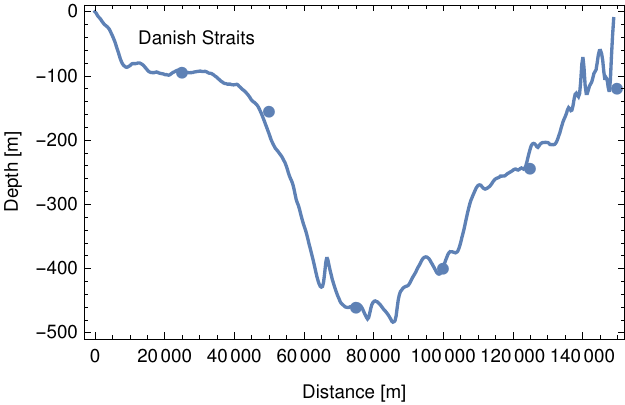}
        \caption{A depth profile of the Danish strait.}
        \label{fig:DanishProfile}
    \end{subfigure}
    \caption{The Danish Strait.}\label{fig:Danish}
\end{figure}

The Danish straits provide an intermediate case between narrow
chokepoints and ocean-scale barriers. Under the same assumptions as in
Sec.~\ref{subsec:Gibraltar}, a 100\,MW reactor can reach about
$2\sigma$ with 10 detectors at 5\,km spacing; reducing to five
detectors at 10\,km spacing lowers significance to about $1\sigma$.

\begin{figure}[h]
    \centering
    \includegraphics[width=\textwidth]{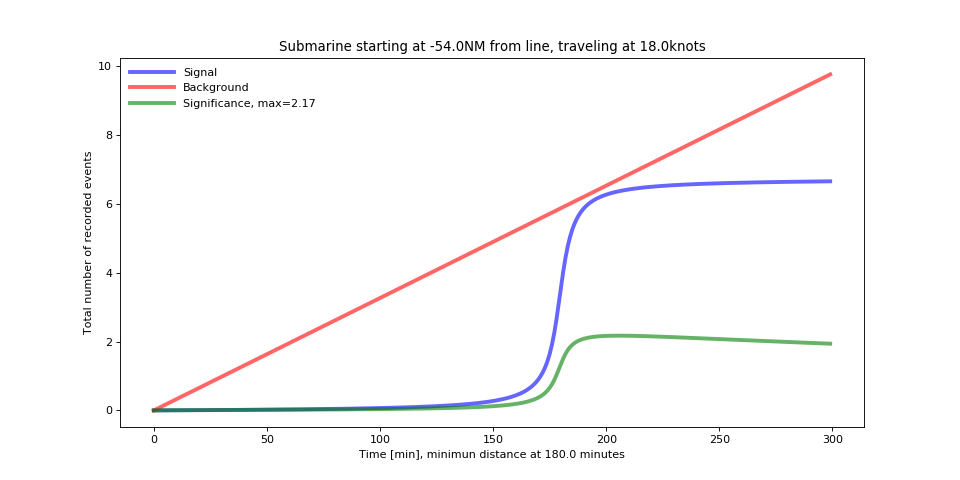}
    \caption{Representative detector-spacing scan for the Danish straits.}
    \label{fig:DanishStudy}
\end{figure}

\subsection{Tsushima strait (Korea-Japan)}
The Tsushima Strait (between Korea and Japan) is another candidate for
a regional barrier. A simplified geometry of roughly 40\,km width and
300\,m depth indicates strong sensitivity to detector spacing and
threshold definition, similarly to the Danish case. In this revision we
therefore do not quote fixed Tsushima module counts in headline claims;
additional work should use detailed local bathymetry, shipping lanes,
and explicit threshold conventions to derive auditable values.

\end{document}